\documentclass[a4paper]{jpconf}
\usepackage{amsmath,amssymb,bbm}
\usepackage{amsfonts}
\usepackage{graphicx}
\usepackage{pgf}
\usepackage{multirow}

\usepackage{ulem}
\usepackage{comment}

\newcommand{\bra}[1]{\langle #1 |} 
\newcommand{\ket}[1]{| #1 \rangle } 
\newcommand{\Schr}{{Schr\"{o}dinger}}
\newcommand{\Ho}{\hat{H}}
\newcommand{\ro}{\hat{\rho}}
\newcommand{\Mc}{\mathcal{M}}

\newcommand\xo{\hat{x}}
\newcommand\xav{\langle\xo\rangle}
\newcommand\psik{|\psi\rangle}

\begin{document}
\title{Causality violation of \Schr--Newton equation: direct  test on the horizon?}

\author{Lajos Di\'osi}

\address{Wigner Research Centre for Physics, Budapest 114, P.O.Box 49, H-1525 Hungary\\
                  E\"otv\"os Lor\'and University,  Budapest, H-1117, P\'{a}zm\'{a}ny P\'{e}ter stny 1/A, Hungary} 
\ead{diosi.lajos@wigner.hu}

\begin{abstract}
We quote a definitive simple proof that neither classical stochastic dynamics nor quantum dynamics
can be nonlinear if we stick to their standard statistical interpretations. A recently
proposed optomechanical test of gravity's classicality versus quantumness is based on
the nonlinear \Schr--Newton equation (SNE) which is the nonrelativistic limit of standard semiclassical gravity.
While in typical cosmological applications of semiclassical gravity the predicted  violation of causality is ignored,
it  cannot be disregarded in applications of the SNE  in high sensitive laboratory tests hoped for the coming years.   
We reveal that, in a recently designed experiment, quantum optical monitoring of massive probes predicts fake 
action-at-a-distance (acausality) on a single probe already.  The proposed experiment might first include the direct
test of this acausality.   
\end{abstract}

Linearity of quantum dynamics hangs on the statistical interpretation of the quantum states.
Also linearity of the classical stochastic dynamics hangs on the statistical interpretation of classical probability distributions.
The following arguments \cite{diosi2007short} constitute the unified proof, valid for the dynamics of both the quantum states $\ro$ or 
the probability distributions $\rho$ provided they are subjects to their standard statistical interpretation.
The ensemble of quantum (or classical) systems belonging to a given $\ro$ (or $\rho$) will be visualized 
by a large number of elements inside an urn.
First, we consider weighted statistical mixing of two states, 
the resulting state must be the weighted sum of the initial two states (fig. 1).
\begin{figure}[h]
\begin{center}
\includegraphics[width=.50\textwidth]{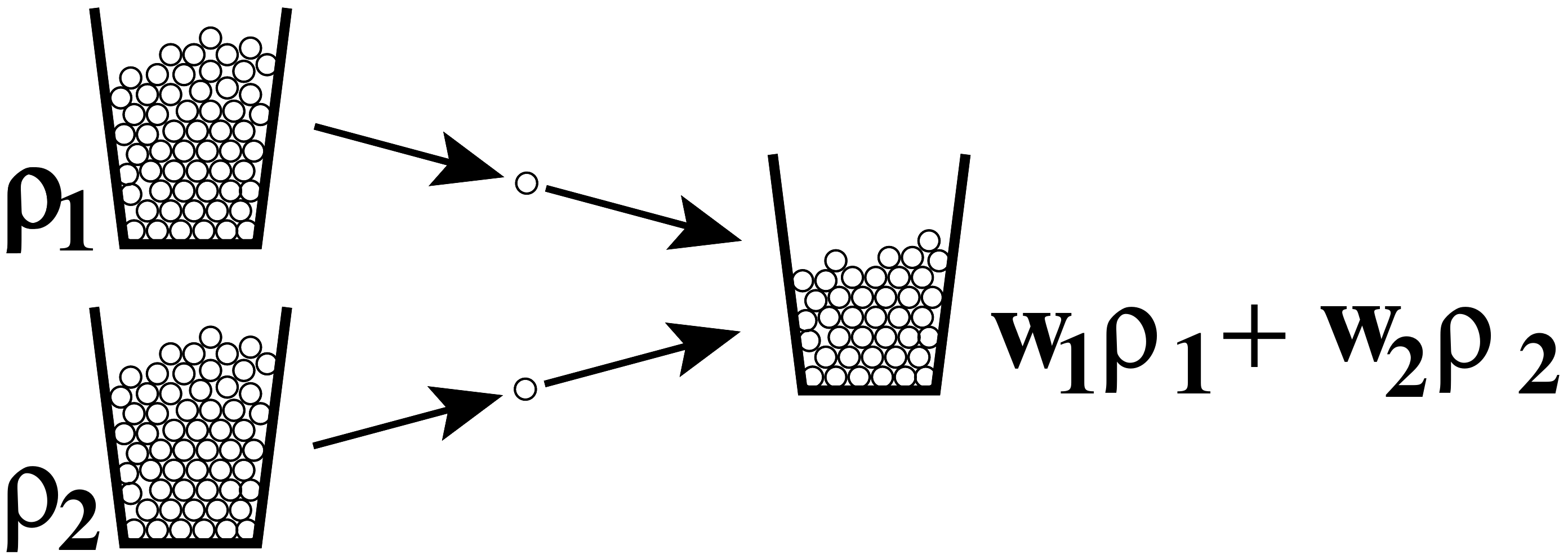}
\caption{Mixing of two states $\rho_1,\rho_2$ in the ensemble interpretation. Random elements from the two urns are
pulled out one by one with probabilies $w_1,w_2$, respectively,  and placed in the third urn which 
represents the state $w_1\rho_1+w_2\rho_2$.}
\end{center}
\end{figure}
Second, we consider a single dynamical step, an operation on the state which is represented 
by a norm-conserving and  (completly) positive map $\Mc$, the resulting state is $\Mc\rho$ (fig. 2).
Note that at this stage the map $\Mc$ is not necessarily linear.
\begin{figure}[h]
\begin{center}
\includegraphics[width=.50\textwidth]{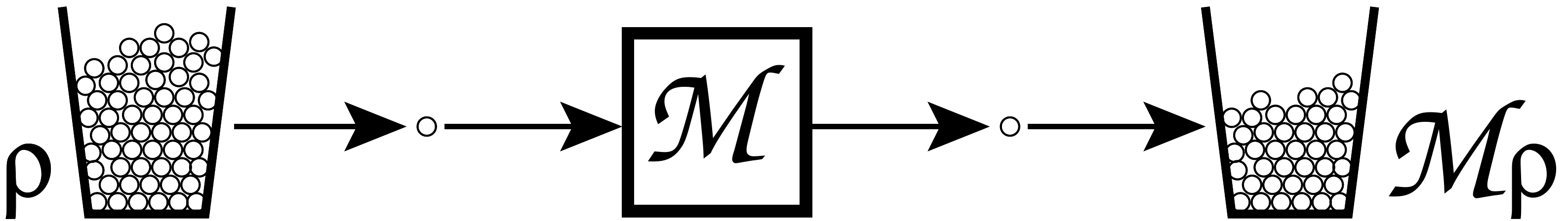}
\caption{Operation in the ensemble interpretation. Randomly chosen elements of the urn on the left one by one undergo the
operation $\Mc$, and go to the urn on the right which represents the state $\Mc\rho$.}
\end{center}
\end{figure}
But really, could the map $\Mc$ be nonlinear?
In one step we get the answer. Consider the combination of the above mixing and operation in 
different orders. Perform operation before mixing, and perform them the other way around (fig. 3).
To keep sense of the statistical interpretation we must assume that
the resulting states do not differ. 
Interchangeability of operation and mixing is a 
mandatory feature of statistical interpretation.
It follows that the resulting states must be identical:
\begin{equation}%1
w_1\Mc\rho_1+w_2\Mc\rho_2=\Mc\left(w_1\rho_1+w_2\rho_2\right).
\end{equation}
This relationship is just the mathematical condition of the linearity of the map $\Mc$.
Nonlinear dynamics of the quantum state $\ro$ (or the classical $\rho$) and the statistical interpretation
of $\ro$ (or  $\rho$) are excluding each other.  
\begin{figure}[h]
\begin{center}
\includegraphics[width=.8\textwidth]{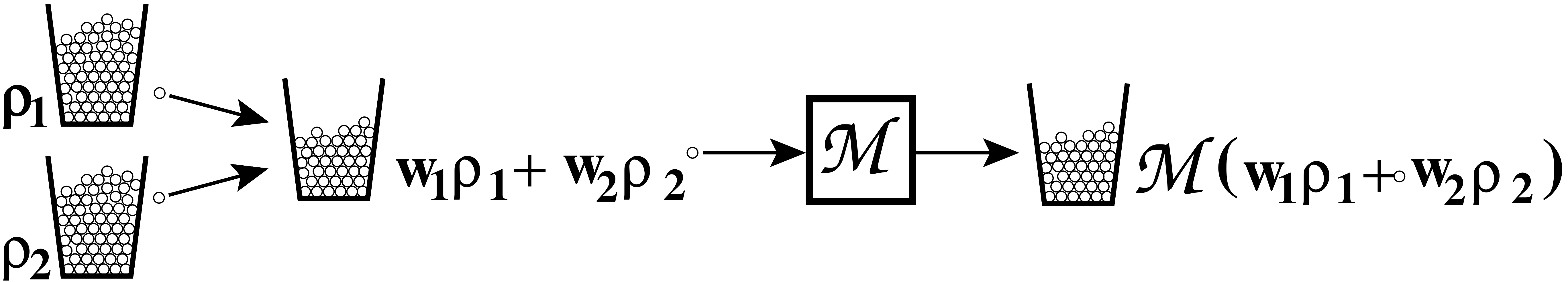}
\vskip12pt
\includegraphics[width=.8\textwidth]{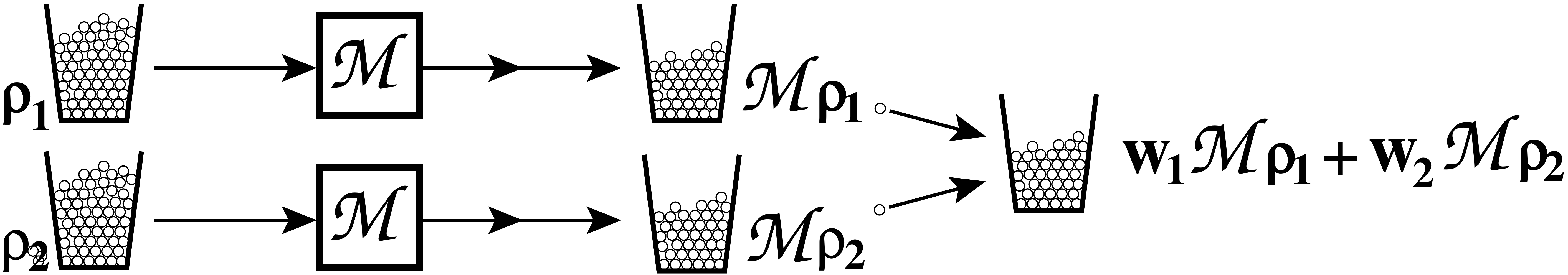}
\caption{Successive combination of mixing and operation, in the ensemble interpretation. 
Mixing is followed by operation, resulting in the state $w_1\Mc\rho_1+w_2\Mc\rho_2$ (top).
Operation is followed by mixing, resulting in the state $\Mc(w_1\rho_1+w_2\rho_2)$ (bottom).}
\end{center}
\end{figure}

Although overshadowed by the conflict with standard statistical interpretation,   
fundamentally nonlinear \Schr~equations are not yet definitively rejected.
The semiclassical theory of gravity is the major example \cite{moller1962, rosenfeld1963}.
The Hamiltonian depends on the wavefunction $\psi$ of the quantized matter because the
classical metric is the solution of the semiclassical Einstein equation: 
\begin{equation}%2
G_{ab}=\frac{8\pi G}{c^4}\bra{\psi}\hat{T}_{ab}\ket{\psi},
\end{equation}
where $G_{ab}$ is the Einstein tensor and $\hat{T}_{ab}$ is the quantized matter energy-momentum tensor. 
In the nonrelativistic limit the semiclassical gravity reduces to the 
the \Schr--Newton equation \cite{diosi1984,penrose1996}. 
The SNE contains the $\psi$-dependent
meanfield Newton potential $\Phi_\mathrm{mf}$ defined by the Newton--Poisson equation 
\begin{equation}%3
\Delta\Phi_\mathrm{mf}(r)=-4\pi G\bra{\psi}\hat{\mu}(r)\ket{\psi},
\end{equation}
where the operator $\hat{\mu}$ stands for the mass density field.  The  SNE in itself can be a legitimate
dynamics of classical gravity and quantized matter just it is not consistent with rhe usual statistical
interpretation of the wavefuction $\psi$.
Standard collapse mechanism is not legitimate for the solutions of the SNE. When, in the lack of 
consistent  alternatives, one combines standard collapse with the SNE then
unaccaptable  features will arise \cite{diosi2016nonlinear}. 
The most spectacular feature is violation of locality by fake action-at-a-distance (superluminality) \cite{gisin1990}.

Semiclassical gravity is indispensable in quantum cosmology. 
The controversies of quantum measurement and collapse remain innocent until the quantum
uncertainties of the energy-momentum around its mean value 
$\bra{\psi}{\hat{T}}_{ab}\ket{\psi}$ are small and can be ignored. Typical cosmological applications belong to this
area of grace where the semiclassical theory works as an effective theory. 
In the Newtonian limit, i.e., in the case of the SNE the situation can be dramatically
different. In the laboratory we may  prepare and control  massive probes with definite positional quantum
uncertainty, meaning definite quantum uncertainty of the  mass distribution around  
$\bra{\psi}\hat{\mu}(r)\ket{\psi}$. A precise interpretation of the wavefunction can no longer be avoided. 
We are going to show that the monster of acausal prediction already comes out in the simplest scenarios .

Consider the quantized position $x$ of a mass $m$. The SNE takes a simple form in the leading order
of the position quantum uncertainty $\xo-\xav$ \cite{diosi2007,yang2013macroscopic}: 
\begin{equation}%4
i\hbar\frac{d\psik}{dt}=\left(\Ho+\tfrac12 m \omega_G^2(\xo-\xav_\psi)^2\right)\psik,
\end{equation}
where $\xav_\psi=\bra{\psi}\xo\ket{\psi}$. The coefficient of the nonlinear self-attraction term 
contains the `Newton oscillator' frequency \cite{diosi2013}, also called \Schr--Newton frequency\footnote
{The notation $\omega_{SN}$ and name \Schr--Newton frequency, widespread after ref. \cite{yang2013macroscopic}, 
are slightly misleading since the frequency is classical Newtonian originally \cite{diosi2013}, 
quantumness is only brought in if the effective density $\varrho$ depends on $\hbar$.}, 
defined by $\omega_G^2=\mathrm{const.}\times G\varrho$ where $\varrho$ is an
effective density of the mass and the constant is a geometric factor. 
Suppose at $t=0$ the probe interacts shortly with a light pulse of incoming state  $\ket{\mbox{light}}$ and after the scattering
they are left in the entangled state  $\ket{\Psi_0}$:
\begin{equation}%5
\psi_0\otimes\ket{\mbox{light}}\Rightarrow\ket{\Psi_0}.
\end{equation}
For $t\rangle0$, the entangled state $\ket{\Psi_t}$ will evolve according to the following SNE:
\begin{equation}%6
i\hbar\frac{d\ket{\Psi}}{dt}=\left(\Ho+\tfrac12 m \omega_G^2(\xo-\xav_\Psi)^2+\Ho_\mathrm{light}\right)\ket{\Psi},
\end{equation}
where $\xav_\Psi=\bra{\Psi}\xo\ket{\Psi}$. Since after  $t=0$  the probe is a closed system, it must have closed
dynamics. Indeed, the probe's reduced density operator $\ro=\tr_\mathrm{light}\ket{\Psi}\bra{\Psi}$ evolves as follows:
\begin{equation}%7
i\hbar\frac{d\ro}{dt}=\left[\Ho+\tfrac12 m \omega_G^2(\xo-\xav_{\ro})^2,\ro\right].
\end{equation}
This nonlinear master equation is exceptional. In general, the SNE equation does not apply to mixed states but
pure ones \cite{diosi2016nonlinear}. But it applies in our special case, i.e.,  after the probe has scattered  a massless system.

So far we assumed that the scattered light remains undetected. Now we consider the same scenario with
the detection of the outgoing light. For simplicity, assume that the detection is performed right after the scattering.
We are trying the old theory of measurement.
Let us measure the outgoing light on $\ket{\Psi_0}$ in a projective measurement. The mixed state of the probe
collapses to a conditional pure state
\begin{equation}%8
\ro_0\Rightarrow\ket{\psi_n}\bra{\psi_n}
\end{equation}
with a given probability $p_n$. For $t\rangle0$ the conditional state $\ket{\psi_n}$ evolve again with the SNE (4).
Right after the detection at $t=0$ the unconditional state of the probe is still the same as it were without the detection: 
\begin{equation}%9
\sum_n  p_n\ket{\psi_n}\bra{\psi_n}=\ro_0,
\end{equation}
as it used to be in standard theory since the detection should not influence the unconditional state of the probe.
It if did  then the detection of light at one place would influence the observables of the probe at a different place.
Exactly this fake action-at-a-distance (acausality) will develop for $t\rangle0$  because the components $\ket{\psi_n}$
on the l.h.s. of eq. (9) evolve by the SNE (4), the r.h.s. evolves by the master eq. (7) and the two sides become different.   
Namely, the unconditional state of the probe will evolve differently if we detect the light or we do not (fig. 4).
\begin{figure}[h]
\begin{center}
\includegraphics[width=.7\textwidth]{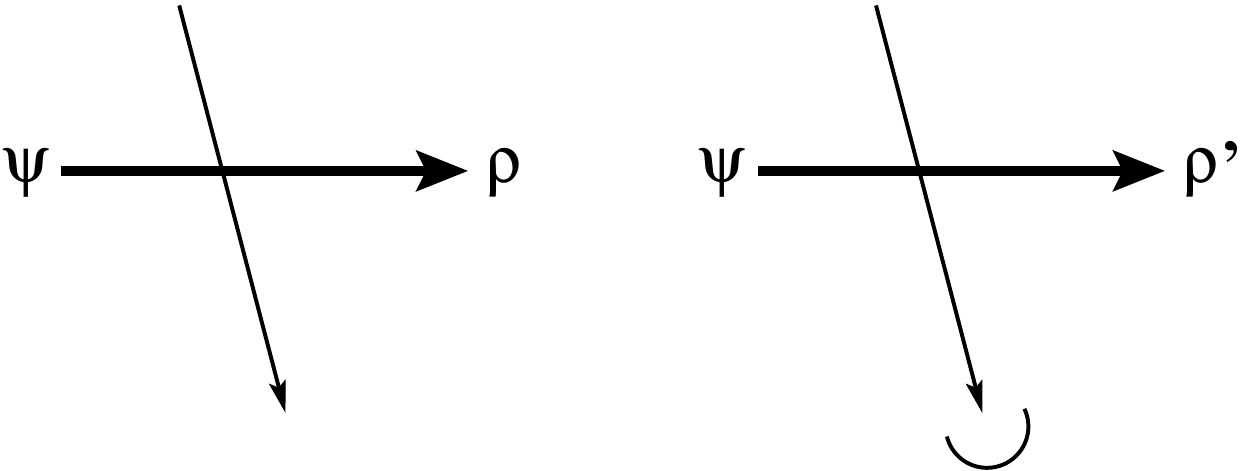}
\end{center}
\caption{The massive probe (fat line) in pure input state $\Psi$ scatters the light pulse (thin line). 
The light is leaving undetected and the reduced state of the probe becomes $\ro$  (left).
If the state of the scattered light is measured,  the unconditional state of the probe becomes $\rho'$,
different from $\rho$ (right).}
\end{figure}

The recently proposed optomechanical experiment 
\cite{liu2023semiclassical,liu2024semiclassical,wilson2024testing}
aims at laboratory tests of  semiclassical gravity versus quantumness of gravity. 
The probes are movable mirrors controlled by laser light
which is monitored  after having been interacted with the mirrors.
The single probe state is assumed to satisfy
the SNE (4)  and an interaction Hamiltonian is responsible for the probe-light coupling.
In the lack of a consistent interpretation of  the nonlinearly evolved
wavefunction, the proposal adopts the standard collapse mechanism of measurement
exactly like we did in eq. (8). 
In a time-discretized picture, one element of  interaction and detection is shown in the right part of fig. 4 
(cf. also in ref. \cite{liu2023semiclassical}) while the left part shows the same without detection.
Since the  post-interaction state
of the probe is different in the two cases it means that the detection activates an action-at-a-distance beyond
our known physics. 

Different conclusions can be drawn from the predicted action-at-a-distance. 
The radical judgement is that, lo and behold, the SNE with the standard collapse mechanism has proved to
be nonsense, in accordance with our proof in fig. 3 and eq. (1). 
On the contrary, one  would keep the design of the experiments   
\cite{liu2023semiclassical,liu2024semiclassical,wilson2024testing} as they are, not caring of the potential
of fake action-at-a-distance. A conclusion in the middle is the interesting one. The toolkits of a monitored 
single mirror, especially under pulsed monitoring  \cite{wilson2024testing}, are already suitable to confirm
or rule out  the predicted action-at-a-distance, so one  should design this test first and shall have to do it 
first when the hoped-for accuracy of optomechanical  control becomes available.

Finally, it is worth noting that the  type of acausality identified here is in fact part of folklore.
If a system, subject to nonlinear dynamics, is entangled with a distant ancilla then the collapse of the ancilla
wavefunction imposes action-at-a-distance on the system which action is developing under the subsequent
nonlinear evolution of the system. Same is true for a classical statistical system under nonlinear dynamics.
If the system is correlated with a distant ancilla then a measurement of the ancilla and the Bayesian update
will activate an immediate action on the system. However, we must recall the truth:  
Collapse or Bayesian update are components of standard statistical interpretation which is fundamentally 
inconsistent with nonlinear dynamics. Yet their heuristic application is fairly instructive.

\section*{Acknowledgements}
This work was supported by the National Research, Development and Innovation Office ``Frontline'' Research Excellence
Program (Grant No. KKP133827) and by EU COST Actions (CA23115, CA23130).

\section*{References}
%\bibliography{diosi2024}
%\bibliographystyle{unsrt}

\end{document}